\documentclass[sigconf, 10pt, authorversion, nonacm]{acmart}

\usepackage{hyperref}
\usepackage[nolist,nohyperlinks]{acronym}
\usepackage[noabbrev, capitalize]{cleveref}
\usepackage[detect-all]{siunitx}
\usepackage[ruled]{algorithm2e}
\usepackage[inline]{enumitem}
\usepackage{subcaption}
\usepackage{pgf}
\usepackage{algorithmic}
\usepackage[english]{babel}
\usepackage{todonotes}
\presetkeys%
    {todonotes}%
    {inline}{}

\newif\ifsubmission{}
\submissiontrue{}

\ifsubmission{}
	\newcommand{\TODO}[1]{}
	\newcommand{\NOTE}[1]{}
\else
	\newcommand{\TODO}[1]{{\color{red}TODO: #1}}
	\newcommand{\NOTE}[1]{{\color{blue}#1}}
\fi

\Crefname{algocf}{Algorithm}{Algorithms}

\usepackage{babel}
\addto\extrasenglish{%

}
\microtypecontext{spacing=nonfrench}

\AtBeginDocument{%
  \providecommand\BibTeX{{%
    \normalfont B\kern-0.5em{\scshape i\kern-0.25em b}\kern-0.8em\TeX}}}

\begin{document}

\title{GWEn --- An Open-Source Wireless Physical-Layer Evaluation Platform}



\author{Alexander Heinrich}
\orcid{0000-0002-1150-1922}
\affiliation{%
  \department[0]{Secure Mobile Networking Lab}
    \institution{Technical University of Darmstadt}
    \city{Darmstadt}
    \country{Germany}
}
\email{aheinrich@seemoo.de}
\authornote{The authors contributed equally to this research.}

\author{Florentin Putz}
\orcid{0000-0003-3122-7315}
\affiliation{%
  \department[0]{Secure Mobile Networking Lab}
    \institution{Technical University of Darmstadt}
    \city{Darmstadt}
    \country{Germany}
}
\email{fputz@seemoo.de}
\authornotemark[1]

\author{S\"oren Krollmann}
\orcid{0009-0000-4711-2622}
\affiliation{%
  \institution{Ambibox GmbH}
    \city{Mainz}\country{Germany}
}
\email{krollmann@ambibox.de}
\authornotemark[1]

\author{Bastian Loss}
\affiliation{%
    \institution{Technical University of Darmstadt}
    \city{Darmstadt}
    \country{Germany}
}
\email{bastian.loss@stud.tu-darmstadt.de}

\author{Waqar Ahmed}
\affiliation{%
  \department[0]{Secure Mobile Networking Lab}
    \institution{Technical University of Darmstadt}
    \city{Darmstadt}
    \country{Germany}
}
\email{wahmed@seemoo.de}

\author{Matthias Hollick}
\orcid{0000-0002-9163-5989}
\affiliation{%
  \department[0]{Secure Mobile Networking Lab}
    \institution{Technical University of Darmstadt}
    \city{Darmstadt}
    \country{Germany}
}
\email{mhollick@seemoo.de}

\renewcommand{\shortauthors}{Heinrich et al.}


\begin{abstract}



Wireless physical layer assessment, such as measuringantenna radiation patterns, is complex and cost-intensive. Researchers often require a stationary setup with antennas surrounding the device under test. There remains a need for more cost-effective and open-source platforms that facilitate such research, particularly in automated testing contexts. This paper introduces the \ac{gwen}, a lightweight multi-axis positioner designed to portably evaluate wireless systems in real-world scenarios with minimal RF interference. We present an evaluation workflow that utilizes GWEn and show how it supports different types of wireless devices and communication systems, including Ultra-wideband, mmWave, and acoustic communication. GWEn is open-source, combining 3D-printed components with off-the-shelf parts, thus allowing researchers globally to replicate, utilize, and adapt the system according to their specific needs.
\end{abstract}



\maketitle


\acresetall

\section{Introduction}
\label{sec:introduction}


\begin{figure}[t]
    \centering
    \includegraphics[width=1.0\linewidth]{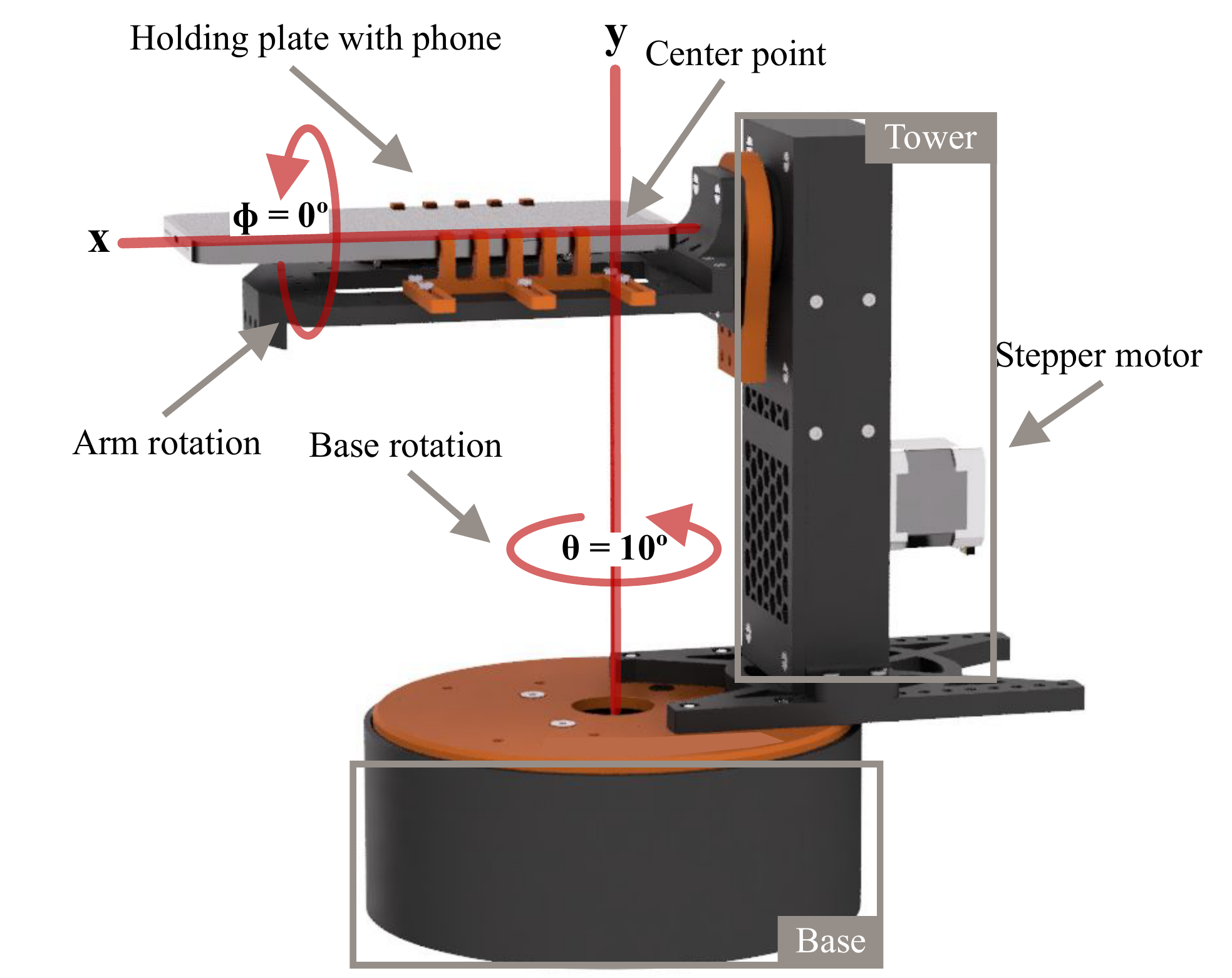}
    \caption{
    3D rendering of an assembled build of our GWEn testbed, highlighting the two rotation axes in red. It can rotate around both axes independently for a full 360° each. The DuT (a smartphone) is placed with its antenna in the center point where both axes meet.
    }\label{fig:gwen_rotation_axes}
\end{figure}

Typical setups for evaluating the wireless physical layer and signal propagation patterns rely on anechoic chambers and extensive antenna arrays, which are costly and cannot mimic real-world conditions~\cite{ieee.1720.antennaMeasurement}. 
While these setups are relevant for the development and initial evaluation of new antennas and the physical layer conditions, researchers strive to evaluate devices in different environments~\cite{finger2022open,picco2011automated,gavrilovich2014test}. 
The wireless sensing community has shown many contributions that could benefit from a portable setup~\cite{Adib2013SeeTW}.

We designed, built, and evaluated the \ac{gwen}, an affordable, open-source, 3D-printed wireless testbed (see \cref{fig:gwen_rotation_axes}). It features a multi-axis positioner, which supports wireless devices such as smartphones, \ac{iot} devices, or \acp{PCB}. \ac{gwen} is compact ($\SI{50}{\centi\meter} \times \SI{50}{\centi\meter}$), lightweight (under \SI{5}{\kilo\gram}), and portable, allowing researchers to study the device's behavior in different environments. Furthermore, researchers can determine antenna radiation patterns which can then be applied in network simulation environments~\cite{eckhoff.antennaPatterns.2016}. We detail several use-cases of the testbed in \cref{sec:UseCases}. 
Complementary to our testbed we define a workflow to perform measurements, guiding researchers to conduct reproducible measurements. 

The contributions of this work are: 
\begin{enumerate}
    \item Design of the testbed \ac{gwen} (\cref{sec:gwen}).
    \item The open-source release of its software and all necessary documentation for 3D printing and manufacturing \ac{gwen}.
    \item A workflow to measure physical layer conditions and antenna radiation patterns with \ac{gwen} (\cref{ssec:gwen_workflow}). 
    \item An evaluation of \ac{gwen} in the context of \ac{uwb} measurements, demonstrating reproducible measurements with minimal influence on the wireless signal (\cref{sec:Evaluation}). 
\end{enumerate}






\section{GWEn}\label{sec:gwen}


\makeatletter
\AC@reset{gwen}
\makeatother

The \ac{gwen} testbed has been designed from the ground up to support evaluations of antenna radiation patterns and signal propagation of wireless devices. This section describes the design, manufacturing process, and interaction with the testbed.

\subsection{Design}

We have designed \ac{gwen} according to the following requirements:

\begin{itemize}
    \item The testbed should allow for reproducible, standardized, and automated measurements, which rotate the \ac{dut} in two axes around a center point.
    \item It should support different types of \acp{dut}, such as smartphones, IoT devices, and \acp{PCB}.
    \item The testbed's design should be 3D printable and openly available so that other researchers worldwide can affordably manufacture it.
    \item The testbed should be lightweight, allowing for mobile measurements at different locations.
    \item The testbed should cause little interference to electromagnetic signals.
\end{itemize}

\begin{figure}[t]
    \centering
    \includegraphics[width=0.7\linewidth]{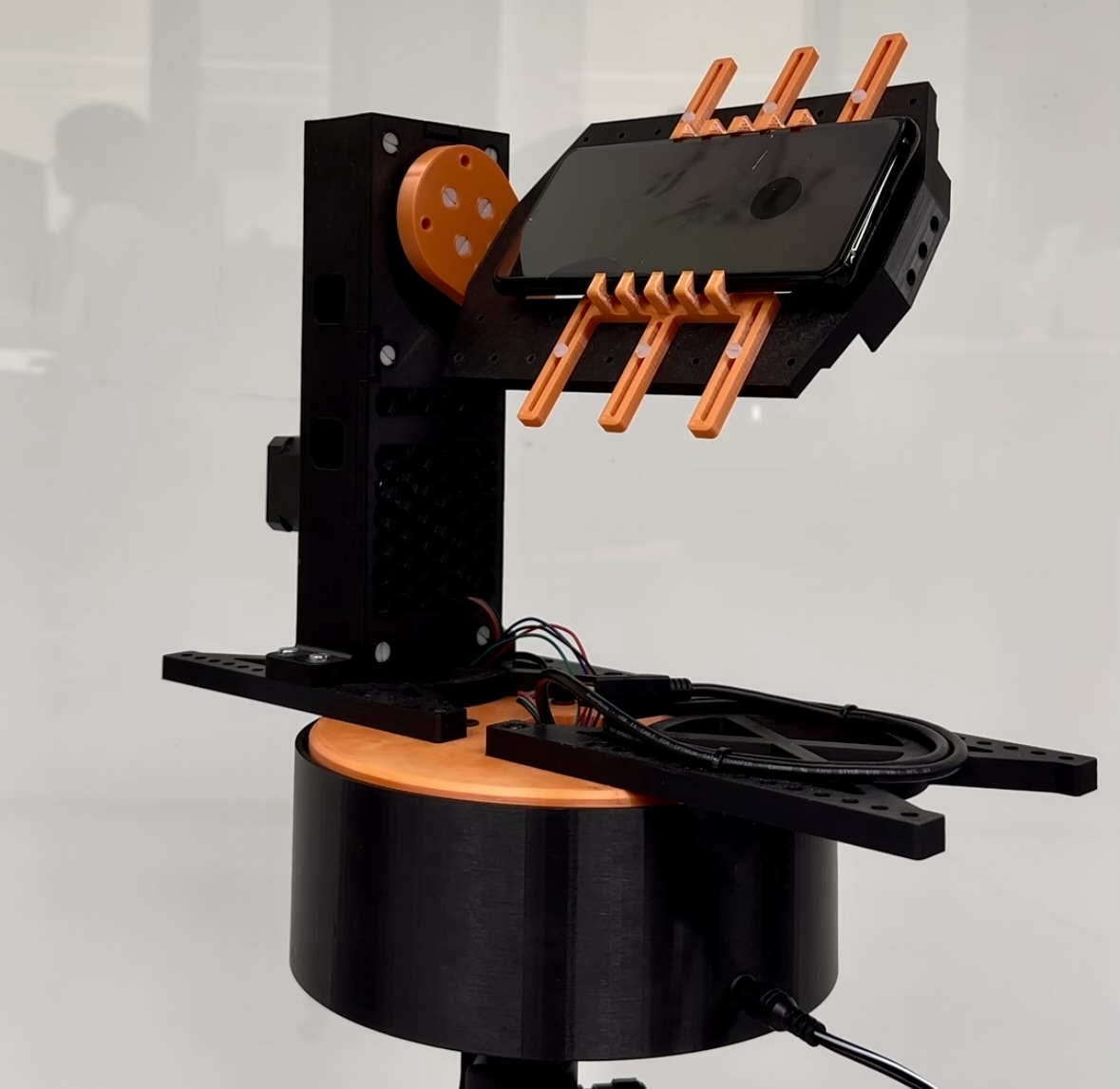}
    \caption{A photo of GWEn with a smartphone mounted.}
    \label{fig:gwen_photo}
\end{figure}

\cref{fig:gwen_rotation_axes} illustrates the \ac{gwen} testbed with two rotation axes characterized by the angles $\phi$ and $\theta$.
The arm, holding the \ac{dut}, rotates around the \textit{x-axis} ($\phi$), and the base rotates around the \textit{y-axis} ($\theta$). Both axes can rotate with a theoretical resolution of \SI{0.01}{\degree}.
To minimize signal interference, we positioned the motors at the bottom of the tower and inside the base. All other electronic components reside in the base. 
\Cref{fig:gwen_photo} shows a photo of a final build of \ac{gwen} with a mounted smartphone. 

In the following, we describe GWEn's main components -- the \textit{base}, the \textit{tower}, the \textit{rotation arm}, and the \textit{linear rail} -- in more detail.

\begin{figure}[t]
    \centering
    \includegraphics[width=1.0\linewidth]{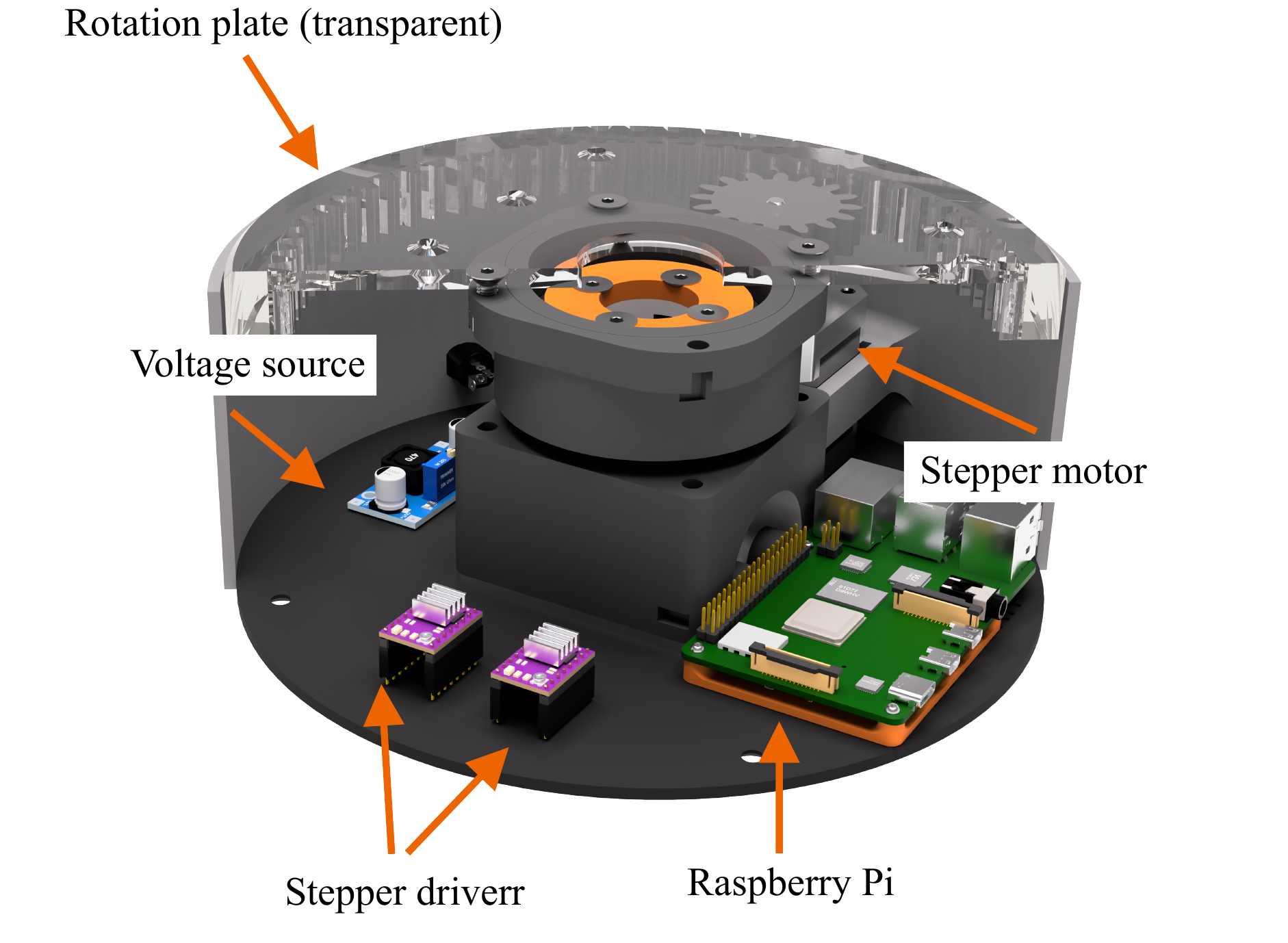}
    \caption{A render of GWEN's Base  containing  power supply, a Raspberry Pi, the motor, and stepper drivers.}
    \label{fig:gwen_base}
\end{figure}

\paragraph{Base} As shown in \cref{fig:gwen_base}, the base contains the electronic parts to control the testbed: the power supply, the stepper drivers for the motors, a Raspberry Pi Model 3B+, and a stepper motor. The Raspberry Pi is the computing module of the testbed, controlling both motors and allowing the  connection of an external device, such as the \ac{dut}, to receive measurement data. The motor in the base is physically connected to the base plate, allowing the entire testbed to rotate around the center of the base. 

\paragraph{Tower} The tower contains a set of belts and ball bearings to transfer the torque from the second motor up to the rotation arm. The tower is attached to the base plate with screws, but one can move it closer or further away from the center, allowing for \acp{dut} of different sizes.

\paragraph{Rotation Arm} The rotation arm is mounted to the tower and rotates around the x-axis. It consists of a holding plate on which researchers mount the \ac{dut} using a width-adjustable clamp system. Researchers can adapt the position of the holding plate to ensure that the antenna of a \ac{dut} is always at the center point.

\paragraph{Linear Rail}
Some experiments also require linear movements of the \ac{dut}. Researchers can mount GWEn on an open-source linear rail system to automate measurements with varying distances.
Another stepper motor, controlled by the Raspberry Pi, moves the linear rail. The linear rail has a theoretical resolution of \SI{0.04}{\milli\meter}~\cite{openbuildsCBeamGantry}.

\subsection{Manufacturing}\label{ssec:gwen_manufacturing}
\ac{gwen} is primarily made of $3$D-printed parts along with standard off-the-shelf components.
The widespread adoption of $3$D printing enables rapid production of required parts, facilitating the precise replication of \ac{gwen} by others. 
In addition, the \ac{pla} filament has low permittivity, allowing electromagnetic waves to pass through with minimal interference~\cite{boussatourDielectricCharacterizationPolylactic2018}. 
We release the $3$D files, a comprehensive list of necessary components, detailed build instructions, wiring diagrams, and the source code in our repository.\footnote{\url{https://zenodo.org/records/7702280}}

Printing all parts for GWEn on a Prusa i3 MK3S+\footnote{\url{https://www.prusa3d.com/category/original-prusa-i3-mk3s/}} takes approximately $4.5$ days.
Assembling \ac{gwen} and soldering the electrical components requires around one week of manual work.
The total material costs for \ac{gwen} amounted to $\sim$300€ in 2024.

\paragraph{Off-The-Shelf Components}
Besides the parts that can be $3$D-printed, our design uses only off-the-shelf parts. These parts should be readily available worldwide, allowing researchers to replicate the build. These parts include stepper motors for device rotation, stepper drivers, a Raspberry Pi, belts, and ball bearings.

\paragraph{Optional Linear Rail}
We have extended GWEn with an open-source linear rail system featuring a C-Beam profile and a standard C-Beam gantry plate from the OpenBuilds project \cite{openbuildsCBeamGantry}.
Mounting GWEn on the linear rail is straightforward, as GWEn has pre-cut holes in the base that precisely align with the gantry plate.

\subsection{User Control}

GWEn is an easy-to-use testbed that requires no programming or technical knowledge to perform testing. This section describes all possible interaction points with the testbed. 
The software of \ac{gwen} consists of three individual parts:
\begin{enumerate*}
    \item~the Python~\textit{web interface},
    \item~the Python~\textit{measurement software},
    \item~the C/C++~\textit{hardware controller}.
\end{enumerate*}

\paragraph{Web Interface}
When turned on, GWEn creates a WiFi access point to which the researcher can connect.\footnote{The access point can be disabled if the 2.4 GHz frequency of WiFi interferes with the frequency that should be evaluated.} At \url{http://gwen.local}, a web interface is presented that enables control of the \ac{dut}'s orientation or to start a test routine.  
Researchers can customize the test routine to collect a defined number of measurements before the device moves on to the next position. By default, the testbed follows \cref{alg:gwen_pseudocode}: 
The testbed moves the \ac{dut} to each possible position by rotating around the x-axis or the y-axis for a defined number of degrees. Researchers can configure the number of degrees that \ac{gwen} moves and achieve to perform fine or coarse measurements.

\begin{algorithm}
\DontPrintSemicolon
\SetAlgoLined
arm\_up = TRUE\;
\While{$\theta$ < 360}{
    gather\_measurements()\;
    \eIf{arm\_up}{
        rotate\_$\phi$\_by(10)\;
    }{
        rotate\_$\phi$\_by(-10);
    }
    \If{(arm\_up AND $\phi$ == 180) OR (!arm\_up AND $\phi$ == 0)}{
        gather\_measurements()\;
        rotate\_$\theta$\_by(10)\;
        arm\_up = !arm\_up\;
    }
}
\caption{Full measurement run with GWEn}
\label{alg:gwen_pseudocode}
\end{algorithm}

\paragraph{Interacting with Python or C}
We wrote the measurement software in Python --- researchers can extend it to their needs.
By default, one interacts with the software using the described web-interface. If more complex movements of the testbed are necessary, researchers can extend the software to implement these. Internally, a C-based hardware controller interacts with the stepper motors to rotate or move the \ac{dut}. Our Python software calls the controller using a command line interface. However,  researchers may use this controller together with their own existing software stack or call it directly using an SSH connection. 

\subsection{Receiving Measurement Data}
\ac{gwen} can receive measurement data from so-called \textit{sources}. A \textit{source} is a customized Python script that fetches data from the \ac{dut} over a USB connection. 
Our software stack already provides a set of sources for UWB-based distance measurements. These sources allow the connection of a UWB-capable \ac{dut}, such as a Google Pixel 8 Pro, to evaluate distance measurements with a separate UWB device while the testbed moves the \ac{dut} to a set of predefined positions (see \cref{sec:usecases:uwb}). 
Adding a new \textit{source} requires a new Python script that defines how to extract measurements from the connected device. Researchers can write new sources for any \ac{dut} that can connect to the testbed.  

\section{Measurement Workflow}\label{ssec:gwen_workflow}


We create a custom workflow to perform antenna radiation and wireless physical layer evaluations. We explain how to mount the \ac{dut}, how to setup the testbed and how to initiate the automated process. 

For an optimal measurement, the antenna of the \ac{dut} must be as close as possible to the center point.
To achieve this with devices of various sizes, researchers can move the \textit{tower} on the \textit{base} outward or inward and the \textit{holding plate} on the \textit{tower} upward or downward.
For most measurements, we require a second device, termed \textit{remote device}, which connects to the \ac{dut} wirelessly. 


We have created a simple and repeatable workflow for measuring with \ac{gwen}:
\begin{enumerate}
    \item Set up \ac{gwen} at the testing location. For best results we recommend to mount \ac{gwen} on a camera tripod. 
    \item Mount the \ac{dut} on the rotation arm. Connect the \ac{dut} via a USB cable to \ac{gwen}. 
    \item Set up the remote device at the desired location and distance. 
    \item If necessary, ensure that the \ac{dut} and remote device are connected wirelessly.
    \item Configure the measurement via the \ac{gwen} web interface and select the correct source to receive measurement data via the USB cable. 
    \item Start the measurement. \Ac{gwen} automatically moves the \ac{dut} to the specified positions, takes the desired number of measurements, and saves them locally.
    \item The measurement runs automatically until it is finished. No monitoring is required during this time. 
    \item Once \ac{gwen} has completed the measurement, download the measurement data using the web interface. 
\end{enumerate}

At the end of each measurement, \ac{gwen} stores a ZIP file, which includes the initial configuration of \ac{gwen} and all measurements performed, ordered by their respective $\theta$ and $\phi$ angles.
Adding extra files, such as a complete system log of the \ac{dut}, is also possible.
In the case of a power outage, \ac{gwen}'s continuous storage of intermediate data enables us to restore data of interrupted measurements. 


\section{Evaluation}\label{sec:Evaluation}

In this section, we assess GWEn regarding two aspects: first, we quantify the signal interference of GWEn's structure on UWB distance measurements. Second, we evaluate whether GWEn allows reproducible measurements by repeatedly performing  identical measurements and comparing the results. 

\subsection{Signal Interference}
\label{sec:gwen_eval:introduced_error}

\Ac{uwb} distance measurements are a good method to evaluate the signal interference of \ac{gwen}. \Ac{uwb} is a pulse-based radio signal which generates signals of very sharp edges, allowing a receiver to accurately determine the ac{toa}. By measuring the round-trip time, \ac{uwb} then allows to calculate the \ac{tof} from which it then computes the distance between two devices. 
If the measured distance changes when using \ac{gwen}, this then quantifies the signal interference. 

\Ac{gwen} comprises several components, including a base, arm, and tower, primarily 3D-printed using \ac{pla}. 
Prior research has found that the \ac{pla} has a low relative permittivity of $2.52$, only resulting in minor signal interferences~\cite{boussatourDielectricCharacterizationPolylactic2018}. 
We can estimate that a solid block of \SI{1}{\centi\meter} \ac{pla} increases the distance by \SI{0.6}{\centi\meter}, based on the relative permittivity~\cite{boussatourDielectricCharacterizationPolylactic2018}. In our manufactured version of \ac{gwen}, the only part which could interfere with signals is the hollow tower.
Furthermore, $3$D printers do not create solid structures but use infill to print durable, lightweight, and cost-saving structures. 
However, some components, such as motors and ball bearings, are metal-based and can influence or block the entire signal. 

\begin{figure}[!t]
  \centering
  \begin{subfigure}{.48\linewidth}
    \centering
    \includegraphics[width=.98\linewidth]{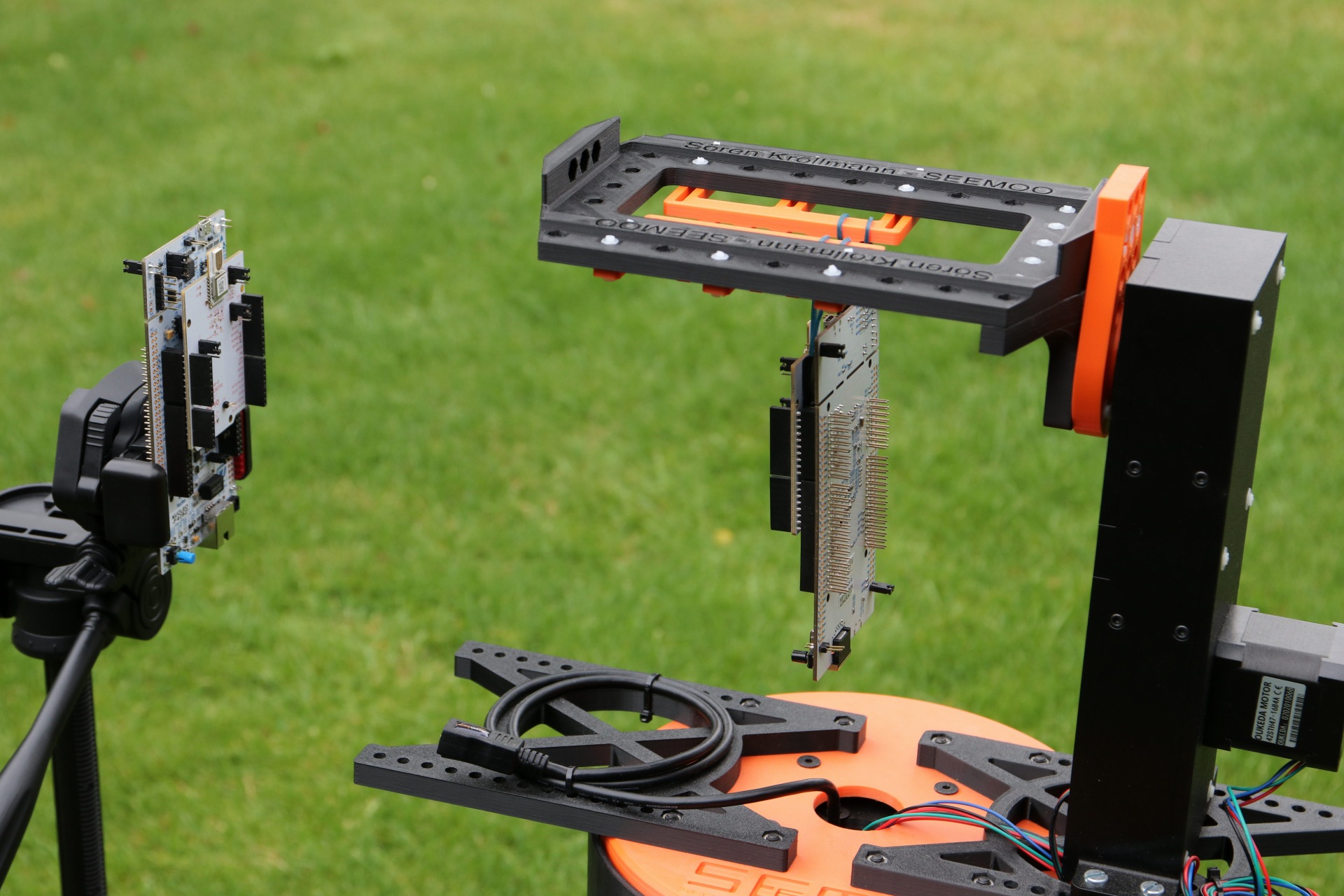}
    \caption{\raggedright{} LOS measurement without obstructions.}\label{fig:t1}
  \end{subfigure}%
  \hfill
  \begin{subfigure}{.48\linewidth}
    \centering
    \includegraphics[width=.98\linewidth]{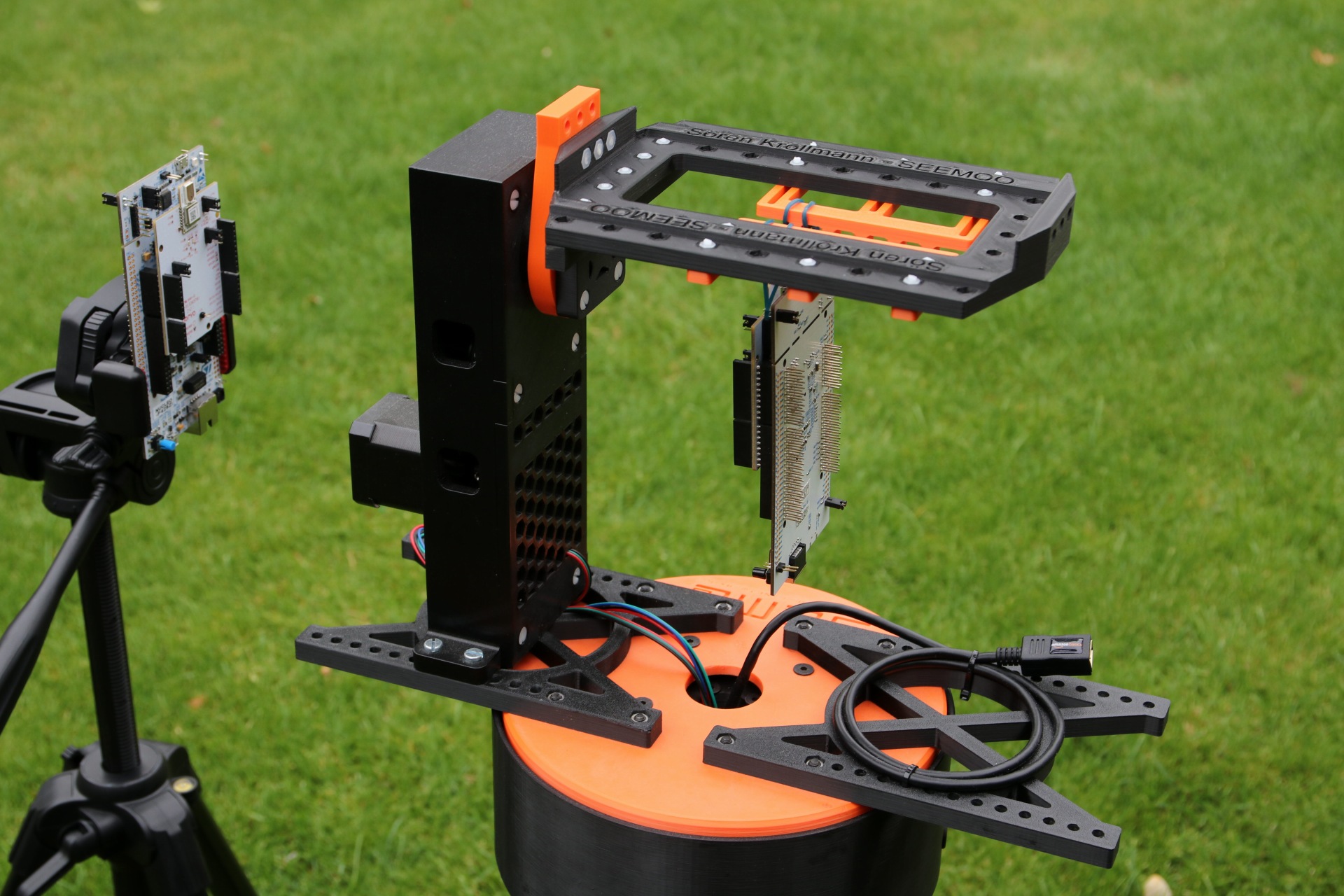}
    \caption{\raggedright{} NLOS measurement with tower between devices.}\label{fig:t2}
  \end{subfigure}
  \caption{Exemplary test setup for evaluating the tower's influence on \ac{gwen}. One UWB transceiver is placed at the same location once with direct \ac{los} and once with \ac{nlos} with the tower in between. 
  }\label{fig:gwen_arm_test}
\end{figure}

In this section, we identify the impact of \ac{gwen} by evaluating the introduced error for the exemplary use case of \ac{uwb} measurements.
To identify the maximum introduced error of \ac{gwen}, we performed two tests using a pair of DWM3000EVB \ac{uwb} development kits. 
First, we placed both devices facing each other with no obstructions in \ac{los} (see \cref{fig:t1}). 
Second, we rotated the base of \ac{gwen} by $\theta=\SI{180}{\degree}$ such that the tower is now directly between both devices (see \cref{fig:t2}). We also adjusted the orientation of the DWM3000EVB to ensure that they had the same orientation as in the first test.  We ensured that the distance between the DWM3000EVB \ac{uwb} devices was equal for both tests, so that the only difference between both setups is the orientation of GWEn.

We performed 10,000 \ac{uwb} measurements in each test. 
To assess the influence of \ac{gwen}, we calculated the mean measured distance and the standard deviation for both measurements.
In \ac{los}, without the tower, we measured a mean distance of \SI{34.660}{\centi\metre} with a standard deviation of \SI{1.345}{\centi\metre}. 
With the tower blocking \ac{los}, the values increased to \SI{36.344}{\centi\metre} and \SI{1.735}{\centi\metre}, respectively.

The tower consists of a motor at the lower end, ball bearings, and two plates of \ac{pla}. 
The tower's presence caused an increase in the average distance by \SI{1.684}{\centi\metre}. 
Also, it caused a rise in the standard deviation by \SI{0.39}{\centi\metre}. 
We can see that the tower of the testbed leads to an increased distance. However, the advertised accuracy of the UWB device is $\pm 10$ cm~\cite{qorvoinc.DW3000DataSheet2020}, and the introduced error is well below that.

\subsection{Reproducible Experiments}
We assess whether experiments with \ac{gwen} can be reproduced by using an experimental setup with two DWM3000EVB boards.
We mounted one of the boards on \ac{gwen} and the other on a camera tripod. We placed them \SI{50}{\centi\meter} apart.
Then, we performed an automated test routine following the workflow in \cref{ssec:gwen_workflow}. 
We recorded $100$ measurements at each position, ensuring enough data for a comparative analysis. 

\begin{table}[t]
  \centering
  \caption{Reproducibility experiments. We performed \# 1--3 in succession, and \# 4 one week later.
  }\label{tab:repdroducibility_experiments}
  \begin{tabular}{lrr}
    \toprule
  \#  & \multicolumn{1}{l}{Mean distance} & \multicolumn{1}{l}{Standard deviation} \\
  \midrule
  1 & 48.867 cm                         & 10.938 cm                              \\
  2 & 48.637 cm                         & 11.174 cm                              \\
  3 & 48.235 cm                         & 11.070 cm                              \\
  4 & 46.603 cm                         & 11.183 cm  \\                           
    \bottomrule
\end{tabular}
\end{table}

\begin{figure}[!t]
  \centering
  \begin{subfigure}[t]{0.45\linewidth}
    \includegraphics[width=\linewidth]{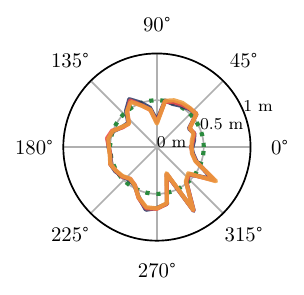}
    \caption{Consecutive Measurements}\label{fig:r1}
  \end{subfigure}
  \begin{subfigure}[t]{0.45\linewidth}
    \centering
    \includegraphics[width=\linewidth]{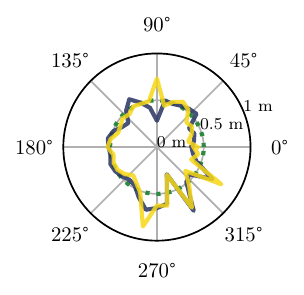}
    \caption{One week apart}\label{fig:r4}
  \end{subfigure}
  \caption{
  The first plot (a) is a comparison of three \SI{360}{\degree} ($\theta$) measurements made directly one after the other without changing the setup. 
  The second plot (b) is a comparison of two \SI{360}{\degree} ($\theta$) measurements made one week apart.
 Arm rotation fixed at $\phi=\SI{90}{\degree}$. The dotted line shows the actual distance (\SI{50}{\centi\meter}). }\label{fig:m_comp_2}
\end{figure}

We performed the measurement series three times in succession. 
\Cref{tab:repdroducibility_experiments} shows the mean distance and standard deviation of our measured data.\footnote{The results cannot be compared with the previous experiment in \cref{sec:gwen_eval:introduced_error}. A different environment and different orientations were evaluated.}
An examination of the first three measurements showed that the values are similar and differ only by a maximum \SI{0.632}{\centi\meter}. 
\Cref{fig:r1} shows each measurement in a plot for a \SI{360}{\degree} base rotation ($\theta$) with a fixed arm angle of $\phi=\SI{90}{\degree}$.
All three graphs have very similar shapes, which applies when comparing an arm angle fixed at a different position.


Finally, we performed a fourth measurement one week later with a fresh setup to evaluate if we could reproduce similar measurements when we set up our testbed again at the identical location (see \cref{tab:repdroducibility_experiments} \#4).
\Cref{fig:r4} shows two measurements, one from the first recording described above and the other from one week later.
The general shape of the two measurements is similar, except for some minor deviations. Environmental changes or slight imperfections could have caused these minor deviations during the setup. 
The mean distance is \SI{2.264}{\centi\meter} shorter than in experiment number one. The standard deviation is also \SI{0.245}{\centi\meter} smaller.

In summary, the results of the first three successive measurements are almost identical. The fourth measurement, which was carried out one week later, showed some minor deviations, but these were still within the manufacturer's accuracy margins. 
In conclusion, our assessment indicates that \ac{gwen} is capable of conducting reproducible measurements.

\section{Use Cases}\label{sec:UseCases}

This section documents our experience evaluating wireless systems for different use cases.
The first physical layer we evaluated is UWB, which we used as the central application example throughout this work. Additionally, we adapted GWEn to facilitate mmWave measurements and measurements of the aerial acoustic physical layer, which can transmit data over sound waves instead of radio signals.

\subsection{Ultra-Wideband}\label{sec:usecases:uwb}

\ac{uwb} is one of the newest physical layers in smartphones. The primary capability of \ac{uwb} is the option to perform short-range distance measurements between two UWB devices of up to 50 m distance with an accuracy between 0.1 m and 0.5~m~\cite{malajnerUWBRangingAccuracy2015, flueratoruHighAccuracyRangingLocalization2022, heinrich2023smartphonesuwbevaluatingaccuracy}. 
These distance measurements are used for security-relevant implementations, such as the digital car key standardized by the car connectivity consortium~\cite{CarConnectivityConsortium}. 
The technology allows one to unlock and start a car based on \ac{uwb} distance measurements from the car to a smartphone. 

Manufacturers like Apple, Google, and Samsung have all released smartphones with \ac{uwb} chips designed for distance measurements. Previous works have already shown that \ac{uwb} is susceptible to physical layer attacks by overshadowing the benign signal, resulting in distance reductions of up to 12~m~\cite{279984}. It is to be discovered if these reductions can also be caused by imperfect placement and signal reflections in real-world environments, putting users of the digital key at risk. 

\begin{figure}
    \centering
    \includegraphics[width=0.9\linewidth]{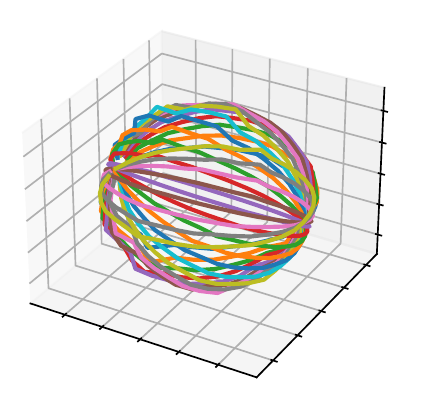}
    \caption{An exemplary measurement with the \ac{gwen} testbed showcasing \ac{uwb} ranging at \SI{5}{\meter} between two DWM3000EVB development kits.}
    \label{fig:gwen_uwb_3d_result}
\end{figure}

We use GWEN for automated evaluations with various smartphones in different positions, environments, and distances. This allows us to identify measurement patterns and potential errors depending on the environment and orientation. 
While our first results are available as a preprint~\cite{heinrich2023smartphonesuwbevaluatingaccuracy}, \cref{fig:gwen_uwb_3d_result} plots the mean measured distance at all positions where \ac{gwen} positions a \ac{dut}. 

\subsection{mmWave}\label{sec:usecases:mmWave}

mmWave Communication utilizes the 30 GHz to 300 GHz band for the 5G and 6G networks \cite{roleofmmwavein5G6G}, resulting in a high throughput, low latency, low interference, and small antenna sizes. As a compromise, mmWave has a reduced range and requires line of sight for communication. The range of communication can be increased by using an array of antennas. The smaller wavelength of mmWave permits packing large antenna arrays in a small area, resulting in narrower beams.
These beams have proven that eavesdropping is more difficult for attackers. Nevertheless, it is still possible if attackers place themselves in the main lobe of the signal. 
We can use GWEn with an integrated transmitter antenna, to change the $\theta$ angle (azimuth) during transmission. Then, we combine the physical positioning of GWEn with directional antenna beams. The main goal of using \ac{gwen} is to characterize \ac{MIMO} antenna beams and to compare radiation patterns of different beamforming algorithms.




\subsection{Acoustic Communication}\label{sec:usecases:acoustic}

\Ac{gwen} not only supports \ac{rf} communication but can also evaluate acoustic communication systems, also known as \textit{data-over-sound}.
Acoustic communication is beneficial for communication between smartphones and IoT devices because it allows the implementation of custom physical layers entirely in software akin to a software-defined radio, utilizing built-in speakers and microphones.
Recent studies have shown acoustic communication to be well-suited for secure device pairing and bootstrapping security applications \cite{scheck2023speakerPairingDemo,shrestha2023soundbased,leinweber2023freespeaker}.
Acoustic communication becomes a practical choice for implementing physical layer security on commercial off-the-shelf devices, where altering the radio physical layer of WiFi or Bluetooth is impossible due to hardware constraints~\cite{zhang2019PriWhisper,putz2020Acoustic,jinKeyAgreementIoT2023}.

Despite its potential for short-range smartphone communication, prior research on acoustic communication only performed limited evaluations, focusing on a small number of device configurations. The directionality of smartphone speakers, affecting communication performance based on device orientation, has been a challenge \cite{zarandyBatNetDataTransmission2020}.
We can use GWEn to address this challenge by performing automated measurements and facilitating the evaluation of different device configurations and orientations.
This evaluation helps determine and subsequently improve the practical suitability of acoustic communication schemes, where smartphones can be oriented in any direction.



\section{Related Work}\label{sec:comparison}

Commercially available multi-axis positioners with high-quality materials and components designed to minimize \ac{rf} interference, promise high precision and accuracy in positioning \cite{etslindgreninc.2112MultiAxisPositioner,diamondengineeringDiamondEngineeringD60502019, diamondengineeringDiamondEngineeringDC4022019,mvg}.
However, these systems come with a significant price tag, ranging from USD $40,000$ to USD $80,000$  as of 2017 \cite{foley2017low}, which might be prohibitively expensive for prototyping or evaluating a smartphone or IoT-based physical layer in research settings. In addition, many of these testbeds are not portable, so they cannot be deployed in real-world scenarios. Another limitation is the lack of software suites to perform automated measurements~\cite{etslindgreninc.2112MultiAxisPositioner, diamondengineeringDiamondEngineeringD60502019, diamondengineeringDiamondEngineeringDC4022019}. 

Besides commercial systems, academia also proposed multi-axis rotation systems for satellite evaluation \cite{gavrilovich2014test} and \ac{rf} antenna measurements \cite{hearn2020open,picco2011automated,rehman2016development}.
~\cite{foley2017low} developed a stationary transmitter and receiver holder for antenna measurements, using acrylic glass to mount the \ac{dut}.
This system, controlled by an open-source \ac{CNC} board, allows rotation in two axes and costs approximately USD $3,000$ in 2017. 
\cite{wagih2021open} introduced a 3D-printed design that rotates the base with an Arduino-controlled robotic arm.
This arm physically tilts the \ac{dut}, leading to a rotation that does not center around a fixed point, affecting measurement accuracy for our use cases.
~\cite{finger2022open} developed a 3D-printed 2-axis rotation system based on a Raspberry Pi.
This design is similar to our system, but GWEn features a complete \ac{pla} enclosure for the gears, belts, and electronic components. 

Compared to the commercial and academic systems mentioned above, GWEn offers several advantages: 

\begin{itemize}
    \item \textbf{Adaptability.} GWEn's mechanical design, an assembly manual, and software are entirely open source. Its plugin architecture allows for easy incorporation of \acp{dut} or the execution of new movement patterns.
    Additionally, our system can be extended with an optional linear rail or even a secondary tower, enabling it to carry loads up to \SI{2}{\kilo\gram}.
    \item \textbf{Portability.} With a weight of less than \SI{5}{\kilo\gram}, GWEn facilitates mobile measurements across diverse environments, whereas the systems mentioned above are typically stationary, with weights ranging from \SIrange{20}{500}{\kilo\gram}.
    \item \textbf{Cost.} GWEn is comparatively cheap to manufacture, with a material cost of $\sim$300€ in 2024.
    \item \textbf{Robust design.} All critical components are housed within the base and tower, making our system suitable for outdoor use.
    \item \textbf{Maintainability.} Users of GWEn can repair it themselves without the need for specialized equipment or tools.  In contrast, damage to commercially available systems often results in expensive and time-consuming repairs, leading to downtime.
    \item \textbf{Efficiency.} GWEn is fully autonomous and includes a user-friendly web interface and scripting capabilities for conducting measurements and managing recorded data. Furthermore, GWEn already supports several \acp{dut}, such as common smartphones and \ac{uwb} devices.
\end{itemize}


\section{Conclusion}\label{sec:conclusion}

In this work, we introduce \ac{gwen}, an open-source testbed that focuses on an easy manufacturing process using 3D-printed and off-the-shelf parts. Its extensible design, demonstrated by adding a linear rail system, enhances adaptability in research environments. 
We successfully manufactured two iterations of the testbed and evaluated their capabilities for reproducible measurements, as well as their minimal influence on wireless signals. 
We showcased three distinct use cases in which \ac{gwen} has already been applied successfully.
In conclusion, \ac{gwen} represents a reproducible and proven testbed offering researchers a powerful open-source tool for the experimental evaluation of wireless communications.

\begin{acronym}
    \acro{uwb}[UWB]{Ultra-Wide Band}
    \acro{hrp}[HRP]{High-Rate Pulse Repetition Frequency}
    \acro{pke}[PKE]{passive keyless Entry}
    \acro{lrp}[LRP]{Low-Rate Pulse Repetition Frequency}
    \acro{twr}[TWR]{two-way ranging}
    \acro{ds-twr}[DS-TWR]{double-sided two-way ranging}
    \acro{ss-twr}[SS-TWR]{single-sided two-way ranging}
    \acro{tdoa}[TDOA]{time difference of arrival}
    \acro{sfd}[SFD]{start-of-frame delimiter}
    \acro{prf}[PRF]{pulse repetition frequency}
    \acroplural{prf}[PRFs]{pulse repetition frequencies}
    \acro{psd}[PSD]{power spectral density}
    \acro{toa}[ToA]{time-of-arrival}
    \acro{tof}[ToF]{time-of-flight}
    \acro{los}[LoS]{line-of-sight}
    \acro{nlos}[NLoS]{non-line-of-sight}
    \acro{sts}[STS]{scrambled timestamp sequence}
    \acro{mitm}[MitM]{Machine-in-the-Middle}
    \acro{ni}[NI]{Nearby Interaction}
    \acro{ble}[BLE]{Bluetooth Low Energy}
    \acro{edlc}[ED/LC]{Early Detect/Late Commit}
    \acro{bpsk}[BPSK]{binary phase shift keying}
    \acro{bpm}[BPM]{burst-position modulation}
    \acro{rssi}[RSSI]{received signal strength indicator}
    \acro{gwen}[GWEn]{\textbf{G}imbal-based platform for \textbf{W}ireless \textbf{E}valuatio\textbf{n}}
    \acro{dut}[DUT]{device under test}
    \acro{ani}[ANI]{Apple Nearby Interaction}
    \acro{fira}[FiRa]{Fine Ranging Consortium}
    \acro{ccc}[CCC]{Car Connectivity Consortium}
    \acro{aoa}[AoA]{angle of arrival}
    \acro{api}[API]{application programming interface}
    \acro{pla}[PLA]{polylactic acid}
    \acro{cdf}[CDF]{cumulative distribution function}
    \acro{mae}[MAE]{mean absolute error}
    \acro{sd}[SD]{standard deviation}
    \acro{rks}[RKS]{remote keyless system}
    \acro{nfc}[NFC]{near-field communication}
    \acro{iot}[IoT]{Internet of Things}
    \acro{prf}[PRF]{pulse repetition frequency}
    \acro{rmse}[RMSE]{root mean squared error}
    \acro{cli}[CLI]{Command Line Interface}
    \acro{rf}[RF]{radio frequency}
    \acro{MIMO}{multiple-input and multiple-output}
    \acro{CNC}{Computerized Numerical Control}
    \acro{PCB}{printed circuit board}
    \acro{SSH}{Secure Shell Protocol}
\end{acronym}


\begin{acks}
 This work has been co-funded by the German Federal Ministry of Education and Research and the Hessian State Ministry for Higher Education, Research, and the Arts within their joint support of the National Research Center for Applied Cybersecurity ATHENE. This work has been co-funded by the LOEWE initiative (Hesse, Germany) within the emergenCITY center [LOEWE/1/12/519/03/05.001(0016)/72].
\end{acks}


\bibliographystyle{ACM-Reference-Format}
\bibliography{main}

\ifsubmission{}
	
\else
	\listoftodos{}
\fi

\end{document}